\documentclass[11pt]{article}

\def\mytitle#1{\setcounter{equation}{0}
\setcounter{footnote}{0}
\begin{flushleft}\Large\textbf{#1}\end{flushleft}
\vspace{0.25cm}}
\def\myname#1{\leftline{{\large #1}}\vspace{-0.13cm}}
\def\myplace#1#2{\small\begin{flushleft}\textit{#1}\\
\texttt{#2}\end{flushleft}}

\def\myclassification#1{\small\noindent
Pacs no :
       #1\vspace{0.5cm}}
\usepackage{graphicx}
\begin{document}

\mytitle{FRW Cosmological model with Modified Chaplygin Gas and Dynamical System}

\vskip0.2cm \myname{Nairwita
Mazumder\footnote{nairwita15@gmail.com}} \vskip0.2cm
\myname{Ritabrata Biswas\footnote{biswas.ritabrata@gmail.com}}
\vskip0.2cm \myname{Subenoy
Chakraborty\footnote{schakraborty@math.jdvu.ac.in}}

\myplace{Department of Mathematics, Jadavpur University,
Kolkata-700 032, India.} { }

\begin{abstract}
The Friedmann-Robertson-Walker(FRW) model with dynamical Dark Energy(DE) in the form of modified Chaplygin gas(MCG) has been investigated. The evolution equations are reduced to an autonomous system on the two dimensional phase plane and it can be interpreted as the motion of the particle in an one dimensional potential.
\end{abstract}
\myclassification{04.20.-q, 04.25.-D,  95.35.+d}
\section{Introduction}
The exciting observational evidences \cite{Riess1,Perlmutter1}in
the last decade raise a challenge to the standard cosmology. To
incorporate the present accelerating phase of the universe within
the framework of Einstein gravity, one has to introduce a
non-gravitating type of matter with a hugely negative pressure (of
the order of its energy density) called dark energy (DE
hereafter). For the mysterious DE, there are only very weak
constrains on its form of an equation of state \cite{Perlmutter1,
Bean1}. The most common candidate for DE is the cosmological
constant $\Lambda$, which can be considered as a perfect fluid
with equation of state (EoS) $p=-\rho$, $\rho=\Lambda$. But it is
discarded due to inconsistency in the value of $\Lambda$ from type
Ia supernovae(SNIa) observations in comparison with the value of
$\Lambda$ interpreted as vacuum energy (i.e., the Planck mass
scale). So usually, investigations in DE are modelled as
quintessence scalar field \cite{Peebles1} or field with barotropic
equation of state \cite{Kamenshchik1}. But the transition from a
universe filled with matter to an exponentially expanding universe
does not necessarily require the presence of a scalar field as the
only alternative. Subsequently, attempts were made to use an
exotic type of fluid-the so-called Chaplygin gas(CG) having the
EoS $p=-\frac{B}{\rho}$ and then it is extended as generalised
CG(GCG) $p=-\frac{B}{\rho^{\alpha}},~B>0,~0\leq\alpha\leq 1$. It
is further generalised to modified CG(MCG) with EoS
\cite{Benaoum1,Debnath}
\begin{equation}\label{1}
p=\gamma \rho-\frac{B}{\rho^{\alpha}}
\end{equation}
with $\gamma,~\beta>0$ and $0<\alpha\leq1$.

This EoS shows a radiation era ($\gamma=\frac{1}{3}$) at one
extreme(when scale factor $a(t)$ is vanishingly small) and a
$\Lambda$CDM model at the other extreme(when $'a'$ is infinitely
large). So at all stages it shows a mixture. Also at an
intermediate stage the pressure vanishes and the matter content is
equivalent to pure dust.

In the present work, we formulate the dynamics of FRW cosmology
with MCG as the matter in it and the evolution equations are shown
to be represented in two dimensional autonomous system by suitable
transformation of variables. The nature of the critical points are
analysed by evaluating the eigen values of the linearized Jacobi
matrix. The paper is organized as follows : Basic equations for
FRW cosmology with MCG are presented and dynamical system is
formulated in section {\bf \ref{chap2}}. Both finite and
asymptotic critical points are analysed in section {\bf
\ref{chap3}}. The paper ends with a short discussion at the end,
in the section {\bf \ref{chap4}}.
\section{Basic Equations for FRW cosmology with MCG} \label{chap2}
The Friedmann equations which governe the dynamics are given by
(assuming $8\pi G=1=c$)
\begin{equation}\label{2}
\frac{\dot{a}^{2}}{a^{2}}+\frac{k}{a^{2}}=\frac{\rho}{3}
\end{equation}
and
\begin{equation}\label{3}
2\frac{\ddot{a}}{a}+\frac{\dot{a}^{2}}{a^{2}}+\frac{k}{a^{2}}=-p
\end{equation}
and the energy conservation relation reads
\begin{equation}\label{4}
\dot{\rho}+3\frac{\dot{a}}{a}\left(\rho+p\right)=0
\end{equation}
From the above field equations the acceleration has the expression
\begin{equation}\label{5}
\frac{\ddot{a}}{a}=-\frac{1}{6}\left(\rho+3p\right)
\end{equation}
Using the MCG EoS in the above conservation equation and integrating we have
\begin{equation}\label{6}
\rho(a)=\left[\frac{1}{\left(\gamma+1\right)}\left(B+\frac{c}{a^{\mu}}\right)\right]^{\frac{1}{\alpha +1}}
\end{equation}
where $\mu=3\left(\gamma+1\right)\left(\alpha+1\right)$ and $c$ is an integration constant.

So the EoS parameter can be writen as
\begin{equation}\label{7}
\omega(a)=\frac{p}{\rho}=\frac{\gamma c-B a^{\mu}}{c+Ba^{\mu}}
\end{equation}
To form an autonomous system from the dynamical equations let us
introduce a new variable
$$x=\dot{a}$$
then we have
\begin{equation}\label{8}
\left.
\begin{array}{c}
\dot{a}=x \\\\
\dot{x}=-\frac{a}{6}\left\{1+3\omega(a)\right\}\rho(a)
\end{array}
\right\}
\end{equation}
Thus set of equations (\ref{8}) form an autonomous system in the
phase plane $(a,~x)$. Further using the phase space variables the
field equation (\ref{2}) can be written as
\begin{equation}\label{9}
\frac{1}{2}x^{2}+V(a)=-\frac{k}{2}
\end{equation}
with
\begin{equation}\label{10}
V(a)=-\frac{a^{2}\rho(a)}{6}.
\end{equation}
Equation (\ref{9}) is termed as the first integral of the solution
of the autonomous system. Also geometrically solutions of the
dynamical system (\ref{8}) should lie on the curve given by
equation (\ref{9}) on the phase plane.
\section{Critical points : Analysis of Dynamical System}\label{chap3}
\begin{figure}
\includegraphics[height=2.6in, width=2.6in]{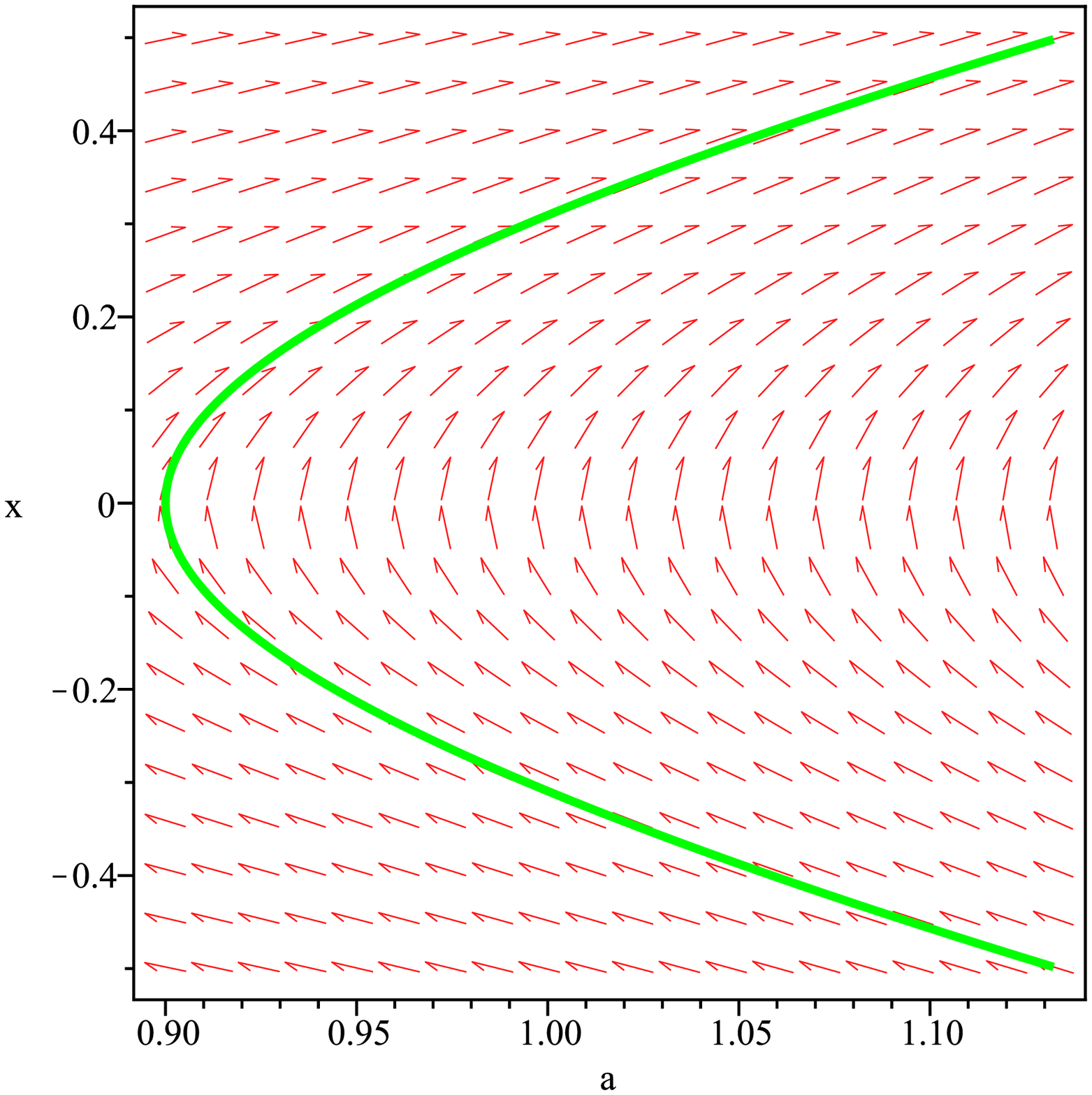}~~
\includegraphics[height=2.6in, width=2.6in]{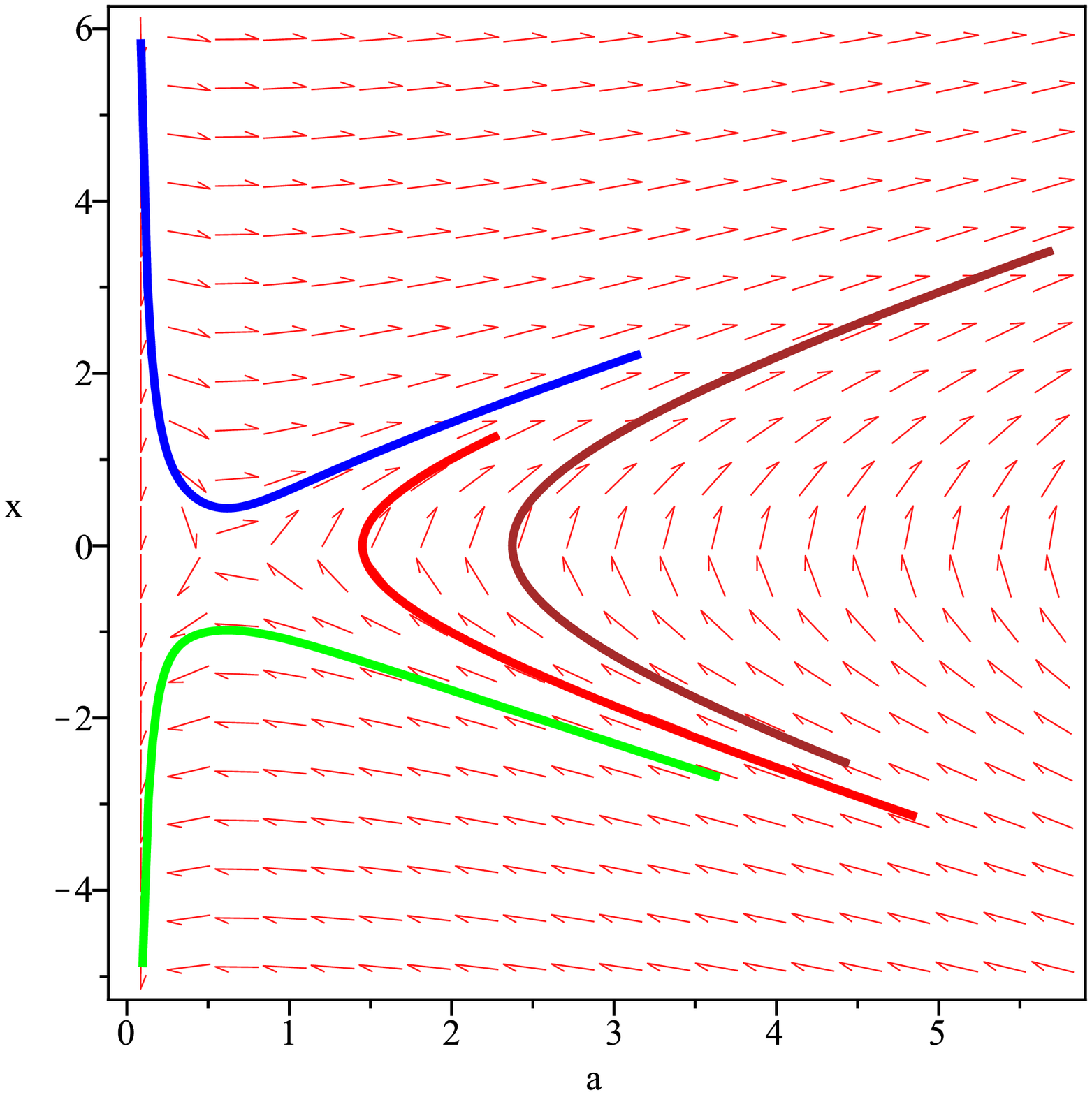}\\
~~~~~~~~~~~~~~~~~~~~~~~~~~~~~~~~~~~~~~~~~~~~~~~~~~~~~~~~~~~~~~~~~Fig.1(a)
~~~~~~~~~~~~~~~~~~~~~~~~~~~~~~~~~~~~~~~~~~~~~~~~~~~~~~~~~~~~~~Fig.1(b)

Fig. 1(a)-1(b) represent the variation of $a-x$ corresponding the
value of the parameters $\alpha := 1.3; B := 3.4; c := 0.7; \gamma
:= 0.3$ \hspace{1cm} \vspace{2cm}

\end{figure}
The critical points on the phase plane have coordinates
\begin{equation}\label{11}
(0,~0)~and ~\left(a_{c}=\left[\frac{c\left(1+3\gamma\right)}{2B}\right]^{\frac{1}{\mu}},0\right)
\end{equation}
and both lie on the $a$-axis. It is to be noted that the critical
points correspond to static universe(s). Moreover, the first one
corresponds to big-bang singularity  and so we will concentrate
mainly on the second critical point. As critical point lies on the
curve (\ref{9}) on the phase plane so we must have
$V(a)=-\frac{k}{2}$ i.e., $k=+1$ is the only possibility. Hence we
can say that the critical points lie in the portion of the phase
plane represented by the trajectories of the closed model.
Moreover, the first integral corresponding to flat model,i.e.,
$\frac{x^{2}}{2}+V(a)=0$ divides the phase plane into two regions
namely $\frac{x^{2}}{2}+V(a)\geq or \leq0$ which correspond to
open and closed models respectively. Further the boundary of the
strong energy condition is the straight line
$a=\left[\frac{c\left(1+3\gamma\right)}{2B}\right]^{\frac{1}{\mu}}$
parallel to the x-axis and the second critical point lies on the
line. The left of this line is the region in the phase space
corresponds to decelerating phase (strong energy condition is
satisfied) while the accelerating phase of evolution (strong
energy condition is violated) is characterized by the right hand
region of the above line \cite{Szydlowski}.

Now we shall examine the character of the second critical point by
evaluating the eigen values of the following linearization matrix
A of the autonomous system :
\begin{equation}\label{12}
A=\left[
\begin{array}{c}
0~~~~~~~~~~~~~~~~~~~~~~~~~~~~~~~~~~~~~~~~~~1 \\\\
\frac{\partial}{\partial a}\left[-\frac{a}{6}\rho(a)\left\{1+3\omega(a)\right\}\right]~~~~~~~~~~~0
\end{array}
\right]_{at~\left(a=a_{c},x=0\right)}.
\end{equation}
So $Tr (A)=0$ and $det(A)=\frac{1}{6}\frac{\partial}{\partial
a}\left[a\rho(a)\left\{1+3\omega(a)\right\}\right]_{\left(a_{c},0\right)}=\frac{1}{2}a_{c}\rho_{a_{c}}\left.\frac{\partial\omega}{\partial
a}\right|_{a=a_{c}}$ with
$\rho(a_{c})=\left(\frac{3B}{1+3\gamma}\right)^{(\frac{1}{\alpha+1})}$
and $\frac{d\omega}{da}=-\frac{B \mu c
a^{\mu-1}\left(1+\gamma\right)}{\left(C+B a^{\mu}\right)^{2}}$.

Then the eigen value problem, namely $det\left[A-\lambda
I\right]=0$, i.e., $\lambda^{2}-\lambda\times Tr(A)+det(A)=0$
simplifies to $\lambda^{2}+det(A)=0$.

Thus we have
\begin{equation}\label{13}
\begin{array}{c}
(i) real~eigen~values~of~opposite~sign~ if~det(A)<0.\\\\
(ii) purely~ imaginary~ conjugate~ eigenvalues~ if~ det(A)>0
\end{array}
\end{equation}
In the first case the critical point is a saddle point while it is
a centre in the other case. As in the present case $det(A)<0$ so
we have critical point which is saddle in nature and hence it is
unstable in character \cite{Perko} as shown in figures 1(a),(b)
for different choices of the parameters.

To complete the study of the dynamical system \cite{Perko} we now
examine the critical points at infinity. To do this we make the
following coordinate transformation in the phase plane to cover a
circle S' at infinity :
\begin{equation}\label{14}
(a,~x)\rightarrow (p,~q): p=\frac{1}{a}, q=\frac{x}{a}; p=0, -\infty<q<+\infty
\end{equation}
Thus the autonomous system (\ref{8}) transfors to
\begin{equation}\label{15}
\left.
\begin{array}{c}
\dot{p}=-pq \\\\
\dot{q}=-\frac{\rho}{6}\left\{1+3\omega\right\}-q^{2}
\end{array}
\right\}
\end{equation}
and the first integral becomes
\begin{equation}\label{16}
q^{2}+2p^{2}V\left(\frac{1}{p}\right)=-kp^{2}
\end{equation}
The critical points on the circle at infinity for the autonomous system (\ref{15}) are
\begin{equation}\label{17}
\left.
\begin{array}{c}
p_{c}=0 \\\\
q_{c}=\pm\sqrt{-\frac{\rho}{6}\left\{1+3\omega\right\}}
\end{array}
\right\}
\end{equation}
which shows that critical points at infinity exists only in the
portion of the phase plane where strong energy condition is
violated(i.e., in the accelerating domain). It should be noted
that for the autonomous system (\ref{15}) there are other critical
points in the finite domain which are already taken in to account.

The linearization matrix which characterizes the nature of the critical points (\ref{17}) is given by
\begin{equation}\label{18}
A=\left[
\begin{array}{c}
-q_{c}~~~~~~~~~~~~~~~~~~~~~~~~~~~~~~~~~~~~~~~~~~0 \\\\
-\frac{1}{6}\frac{d}{dp}\left\{\rho\left(1+3\omega\right)\right\}~~~~~~~~~~~~~~~~~~~-2q_{c}
\end{array}
\right].
\end{equation}
So clearly, $Tr(A)=-3q_{c}$ and let $det(A)=2q_{c}^{2}$ and the
eigen values are $-q_{c}$ and $-2q_{c}$,i.e., eigenvalues are real
and of same sign. So both the critical points are node and they
are asymptotically stable when $q_{c}>0$ and are unstable for
$q_{c}<0$

\section{Discussion}\label{chap4}
In this work we study the FRW comology with MCG as the matter
contained. The evolution equations are reduced into an autonomous
dynamical system with a suitable change of variables. The
dynamical system has a first integral and solutions to the system
should lie on the curve represented by the first integral in the
phase plane. Also the first integral can be interpreted as the
energy conservation relation of a particle moving in an
one-dimensional potential. Except the big bang singularity, there
is only one critical point for the system which is saddle type and
hence unstable in nature. Also there are two critical points on
the circle at infinity and are node type asymptotically stable or
unstable depending on the sign of the coordinate of the critical
point. Finally, it should be noted that strong energy condition
has a vital role for existence and type of critical points both at
finite domain and at infinity.\\\\

{\bf Acknowledgement :\\}
RB and NM are thakful to State Govt. of West Bengal, India and CSIR, India respectively. All the authors are thankful to IUCAA, Pune as this work was done there during a visit.

\end{document}